\documentstyle[aps,preprint]{revtex}
\begin{document}
\draft

\title
{
Near critical states
of random Dirac fermions
}
\author{Y. Morita and Y. Hatsugai}
\date {May 2}
\address
{
Department of Applied Physics, University of Tokyo,
7-3-1 Hongo Bunkyo-ku, Tokyo 113, Japan
}
\maketitle
\begin{abstract}
Random Dirac fermions
in a two-dimensional space
are studied numerically.
We realize them
on a square lattice
using the $\pi$-flux model with
random hopping.
The system preserves two symmetries,
the time-reversal symmetry and
the symmetry denoted by ${\{}{\cal H},{\gamma}{\}}=0$
with a $4\times 4$ matrix $\gamma $
in an effective field theory.
Although it belongs to the orthogonal ensemble,
the zero-energy states
do not localize
and become {\it critical}.
The density of states vanishes at zero energy
as $\sim E^{\alpha}$ and
the exponent $\alpha$ changes
with strength of the randomness,
which implies the existence of the critical line.
Rapid growth of the localization length
near zero energy is suggested and
the eigenstates near zero energy
exhibit anomalous behaviour
which can be interpreted as a {\it critical slowing down}
in the available finite-size system.
The level-spacing distributions close to zero energy
deviate from both the Wigner surmise and the Poissonian,
and exhibit critical behaviour
which reflects the existence of critical states at zero energy.
\end{abstract}
\narrowtext
Dirac fermions often appear
in condensed matter physics, for example,
a transition
between different quantum Hall states {\cite {wu1,wu2,nnn}},
two-dimensional graphite sheets {\cite {graphite}},
a mean-field theory of the t-J model ('$\pi$-flux state') {\cite {pi-flux}}
and d-wave superconductors {\cite {d-wave}}.
It is then natural to investigate what happens
when disorder is included.
Random Dirac fermions
in a two-dimensional space
were investigated by several groups
{\cite {d-wave,ludwig,lan,site1,site2,dos1,dos2}}.
Possible appearance of
non-localized states, critical states,
in random Dirac fermions
was pointed out in {\cite {ludwig}}.
Recently, this disordered critical state was realized
in a lattice model,
where
it was crucial
to preserve
a symmetry
denoted by ${\{}{\cal H},{\gamma}{\}}=0$
with a $4\times 4$ matrix $\gamma $
in an effective field theory
{\cite {lan}} (see below).

In this paper,
we study the random Dirac fermions numerically beyond the zero modes.
In order to realize
the massless Dirac fermions on a two-dimensional lattice,
we use a tight-binding model on a square lattice
with half a flux quantum ('$\pi$ flux') per plaquette,
which is described by the Hamiltonian
\begin{equation}
H_{\it pure}=\sum _{<i,j>}c_{i}^{\dagger}t_{ij}c_{j}+h.c,
\label{ham}
\end{equation}
where the summation is over the nearest-neighbor bonds.
The hopping matrix elements are given by
$t_{j+{\hat x},j}=(-1)^{j_{y}}$
and
$t_{j+{\hat y},j}=1$,
where $j=(j_{x},j_{y})$, ${\hat x}=(1,0)$ and ${\hat y}=(0,1)$.
In the momentum space,
the Hamiltonian is rewritten as
\begin{eqnarray}
H_{\it pure}=
2
\sum _{(k_{x},k_{y})}
\psi _{(k_{x},k_{y})}^{\dagger}
\left(\begin{array}{cc}
\cos k_{y}&\cos k_{x}\\
\cos k_{x}&-\cos k_{y}\\
\end{array}\right)
\psi _{(k_{x},k_{y})},
\nonumber
\\
\label{kspace}
\end{eqnarray}
where
the summation is over the magnetic Brillouin zone
$[-{\pi},{\pi})\times[0,\pi)$
and
$\psi _{(k_{x},k_{y})}^{\dagger}=
(c^{\dagger }_{(k_{x},k_{y})}, c^{\dagger }_{(k_{x},k_{y}+\pi)}$).
There are two energy bands
$E({\bf k})=\pm 2\sqrt {{\cos}^{2}k_{x}+{\cos}^{2}k_{y}}$
on the magnetic Brillouin zone.
They touch at two momenta,
${\bf k}^{1}=(k_{x}^{1},k_{y}^{1})=({\pi}/2,{\pi}/2)$
and
${\bf k}^{2}=(k_{x}^{2},k_{y}^{2})=(-{\pi}/2,{\pi}/2)$.
Near the degeneracies ${\bf k}^{i}\ (i=1,2)$,
they
behave as
$E({\bf k})\approx
\pm 2\sqrt {(k_{x}-k_{x}^{i})^{2}+(k_{y}-k_{y}^{i})^{2}}\ (i=1,2)$.
Define continuum variables
$\Psi ^{\dagger}(x,y)
=
(
{\psi _{1}}^{\dagger}(x,y),
{\psi _{2}}^{\dagger}(x,y),
{\psi _{3}}^{\dagger}(x,y),
{\psi _{4}}^{\dagger}(x,y)
)
$
by
$c_{j}\sim
a
[\ i^{j_{x}+j_{y}}{\psi _{1}}(x,y)
+i^{j_{x}-j_{y}}{\psi _{2}}(x,y)
+i^{-j_{x}+j_{y}}{\psi _{3}}(x,y)
+i^{-j_{x}-j_{y}}{\psi _{4}}(x,y)\ ]$,
where
$a$ is the lattice spacing
and
$x=aj_{x},y=aj_{y}$.
Then the Hamiltonian becomes in the continuum limit
($a\rightarrow 0$)
\begin{eqnarray}
{\cal H}_{pure}=
2i\int d{\bf x}\
\Psi ^{\dagger}(x,y)
{\Bigg [}
\left(\begin{array}{cc}
\sigma _{1}&0\\
0&-\sigma _{1}\\
\end{array}\right)
{\partial }_{x}
+
\left(\begin{array}{cc}
\sigma _{3}&0\\
0&\sigma _{3}\\
\end{array}\right)
{\partial }_{y}
{\Bigg ]}
\Psi (x,y).
\label{dirac}
\end{eqnarray}
Therefore
our lattice model includes
doubled massless Dirac fermions.

There are several subtleties for the massless Dirac fermions.
When the Fermi energy lies at zero energy, that is,
all the negative energy eigenstates are filled,
the Hall conductivity $\sigma _{xy}$ is ill defined.
An infinitesimal mass determines the $\sigma _{xy}$
in the continuum theory {\cite {diracsigma}}.
Similar phenomenon also occurs in a lattice model
where
an infinitesimal next-nearest-neighbor hopping $t'$ opens a gap
and the $\sigma _{xy}$ is given by $t'/|t'|$ \cite{nnn}.
Therefore the massless Dirac fermions are
at a quantum phase-transition point between differnt quantum Hall states.

Let us consider the effect of randomness
in the hopping matrix elements.
We set
$t_{j+{\hat x},j}=(-)^{j_{y}}+\delta t_{j+{\hat x},j}$
and
$t_{j+{\hat y},j}=1+\delta t_{j+{\hat y},j}$,
where
$\delta t_{j+{\hat x},j}$ and $\delta t_{j+{\hat y},j}$
are random variables
and taken at random with constant probability from $[-W,W]$.
It should be noted
that this model preserves the time-reversal symmetry
and belongs to the orthogonal ensemble.
Another example of the orthogonal ensemble,
Dirac fermions with diagonal disorder,
was studied in \cite {site1,site2}
and it was suggested that
all the eigenstates localize,
which is consistent
with the scaling theory of the Anderson localization\cite {aalr}.
In the case of the random-hopping model,
however,
it was found that the zero mode
does not localize
but become critical \cite{lan}.
Similar phenomenon was found
at the band center
of the quantum Hall states \cite{qhe}.
In the quantum Hall states, however,
the time-reversal symmetry is broken
and
the system belongs to
a different universality class,
the unitary ensemble.
In \cite{lan},
it was confirmed that
parameters for
the critical states
form a critical line
in the parameter space of the Hamiltonian,
which is connected to
pure massless Dirac fermions at zero energy.
We consider
that
the stability of the zero modes
against random hopping matrix elements
is due to a symmetry of our Hamiltonian.
The random hopping matrix elements preserves
the symmetry
in contrast to the diagonal disorder.
In the language of the lattice Hamiltonian ({\ref {ham}}),
the symmetry means that
the transformation $c_{j}\rightarrow (-1)^{j_{x}+j_{y}}c_{j}$
induces sign change of the Hamiltonian.
Thus
the eigenstates always appear in pairs
with energies $E$ and $-E$.
The corresponding transformation
in the continuum Hamiltonian ({\ref {dirac}})
is given by
${\cal H}_{pure}
{\rightarrow}
{\gamma}^{\dagger}{\cal H}_{pure}{\gamma}
=-{\cal H}_{pure} $,
where
${\gamma}={\sigma _{1}}\otimes{\sigma _{1}}$.
Since the random hopping matrix elements preserve
the symmetry,
the continuum Hamiltonian ${\cal H}$
for the random-hopping model
also satisfies
${\{}{\cal H},{\gamma}{\}}
={\cal H}{\gamma}+{\gamma}{\cal H}=0$.
Thus,
taking the lowest order in derivatives,
we obtain the following form
as a possible Hamiltonian
for the effective field theory
\begin{eqnarray}
{\cal H}=
2i\int d{\bf x}\
\Psi ^{\dagger}(x,y)
{\Bigg [}
\left(\begin{array}{cc}
\sigma _{1}&0\\
0&-\sigma _{1}\\
\end{array}\right)
{\partial }_{x}
+
\left(\begin{array}{cc}
\sigma _{3}&0\\
0&\sigma _{3}\\
\end{array}\right)
{\partial }_{y}
+
\sum _{i=1}^{4}a_{i}(x,y)\ {\gamma}^{i}
{\Bigg ]}
\Psi (x,y),
\label{randomdirac}
\end{eqnarray}
where
${\gamma}^{1}={\sigma }_{2}\otimes I$,
${\gamma}^{2}={\sigma }_{1}\otimes {\sigma }_{2}$,
${\gamma}^{3}=-{\sigma }_{2}\otimes {\sigma }_{1}$,
${\gamma}^{4}=I\otimes {\sigma }_{2}$
and $a_{i}(x,y)$ (i=$1,{\cdots},4$) are random variables.

In this paper,
we study random Dirac fermions
numerically
beyond the zero modes.
We diagonalize the Hamiltonian
for finite squares of size
$L^{2}=20^{2}$, $30^{2}$, $40^{2}$ and $50^{2}$.
To obtain reliable statistics,
ensemble average over
$16000$, $16000$, $8000$ and $3360$ realizations is performed
respectively.
The observables
are
density of states $\rho(E)$,
the dimensionless conductance (Thouless number) $g(E)$
and
the level-spacing distribution $P(s)$.

Let us first discuss
the density of states
$\rho(E)=1/L^{2}\sum _{i}\delta(E-E_{i})$.
When there is no randomness,
the $\rho(E)$ vanishes linearly at zero energy.
Recently,
whether the density of states is finite or not
at zero energy
for random Dirac fermions
is controversial
\cite{dos1,dos2}.
The $\rho(E)$'s for different
strength of the randomness are shown in Fig. 1.
We have fitted the data by the power-law form
$\rho(E)=CE^{\alpha (W)}$.
Within the numerical accuracy,
our results support
the vanishing density of states at zero energy
with an anomalous exponent $\alpha (W)$,
which depends on strength of the randomness.

Next, in order to reveal nature of the eigenstates,
let us consider
the dimensionless conductance (Thouless number) $g(E)$.
The $g(E)$
is defined by
$g(E)={V(E)}/{\Delta(E)}$,
where $V(E)$ is an energy shift
obtained by replacing
periodic boundary condition
by
antiperiodic boundary condition
and $\Delta(E)$ is a local mean level spacing
near the energy $E$.
Numerical results for the $g(E)$
are shown in Fig. 2 with $L=$30, 40 and 50, where
ensemble average is performed
within an energy window
whose center is located at each data point.
Rapid enhancement of the $g(E)$
near zero energy is observed in Fig. 2.
It suggests that
the localization length grows rapidly
near zero enrergy.
This is consistent
with the existence
of crtitical states at zero energy.
One may consider
that the zero modes are just on the critical point.
Then one of the possible scenarios is that
the non-zero energy eigenstates are all off critical
and therefore localized.
It suggests an exponential dependence
of the finite-size dimensionless conductance $g(E,L)$.
The $g(E,L)$ obtained numerically
decreases when the system size increases.
It is, however, far from the exponential dependence.
In Fig. 3,
we have plotted the $g(E,L)$ as a function of $1/L$.
It suggests a power-law form $g(E,L)\propto 1/L^{\gamma}$
rather than an exponential form $g(E,L)\propto \exp(-L/{\xi})$.
Although
we can not exclude
a possible existence
of critical states in a finite energy region,
we consider
that
the non-zero energy states may be localized
in an infinite-size system
and
a crossover
from the power-law form to an exponential form
occurs
when the system size increases beyond the localization length.
The localization length of the eigenstates near zero energy
may be large compared to the available system sizes
and
we may say that
the power-law dependence of the $g(E,L)$
is a critical slowing down
in the available finite-size system.
This also suggests
the existence of the critical state at zero energy.

We have also obtained
the level-spacing distribution $P(s)$.
The $P(s)$'s
of the normalized energy separation $s=|E_{n}-E_{n+1}|/{\Delta}(E_{n})$
are shown in Fig. 3 and 4,
where
$E_{n}$ and $E_{n+1}$ are two successive eigenenergies.
The $P(s)$
is well described
by the Wigner surmise $P(s)=As^{\beta}{\exp}(-Bs^{2})$
in the metallic regime
and
becomes the Poissonian $P(s)={\exp}(-s)$
in the insulating regime.
The parameter $\beta$ in the Wigner surmise
reflects the symmetry of the Hamiltonian
and
$\beta$=1, 2 and 4
for
the orthogonal ensemble,
the unitary ensemble
and
the symplectic ensemble
respectively.
The parameters $A$ and $B$ are determined by
$\int _{0}^{\infty} ds\ P(s)=1$ and  $\int _{0}^{\infty} ds\ sP(s)=1$
and, in paticular,
$A={\pi}/2$ and $B={\pi}/4$ for the orthogonal ensemble.
The $P(s)$ characterizes nature of the eigenstates.
States
localized in different spatial regions
are allowed to lie at the same energy.
It means that
the energy levels of the localized states
distribute independently,
which is described by the Poissonian.
On the other hand,
in metals where the eigenstates are extended,
two adjacent energy levels interact strongly,
which brings strong energy repulsion and
$P(s)\sim s^{\beta }$
near $s=0$,
where ${\beta }$ is determined by
the symmetry of the Hamiltonian.
The $P(s)$ is well explained
by a $2\times 2$ random matrix model,
which is the Wigner surmise.
Level statistics
near the mobilty edge
has been studied recently and
the appearance of critical level statistics
is discussed \cite{critical}.
Several numerical studies on level statistics
near the mobility edge
were performed
for e.g.
the three-dimensional Anderson model \cite{3d}
and
the band center of the quantum Hall states \cite{qhe-ls},
which
belong
to
the orthogonal ensemble
and
the unitary ensemble respectively, and
it was found that
the $P(s)$'s deviate
from both the Wigner surmise and the Poissonian
and exhibit critical behaviour, i.e.,
the $P(s)$'s  do follow the Wigner surmise
for small $s$
and
they then deviate from it at higher values of $s$
and show stretched exponential decay.
The $P(s)$'s near zero energy for our model are shown in Fig. 3,
where
the energy window is set $[0.1,0.5]$ \cite{iiwake}.
The location of the energy window
is set sufficiently close to zero energy
compared to the band width,
which corresponds to the energy cut off in the continuum theory.
Thus
we consider that
the system is described
by random Dirac fermions.
We confirmed that the $P(s)$'s near zero energy
do not seriously depend on
the system size and the location of the energy window.
We consider that this is due to the fact that,
since
the location of the energy window
is set near zero energy
compared to strength of the randomness,
the localization lengths are so long
that it exceeds the system size.
Thus,
although the eigenstates may be localized in an infinite-size system,
they behave as critical wavefunctions in a finite-size system.
In fact, the $P(s)$ in Fig. 3
deviate
from both the Wigner surmise and the Poissonian,
and exhibit critical behaviour, i.e.,
the $P(s)$'s  do follow the Wigner surmise
for small $s$
and
they then deviate from it at higher values of $s$
and show stretched exponential decay.
On the other hand,
when
the location of energy window
is set in other regions,
non-critical behaviour is found (see Fig. 4).
For example,
when $L=20$, $W=0.5$ and the center of the energy window is set $E=1.5$,
the randomness is so weak
that the deviation from pure massless Dirac fermion is small
and the $P(s)$ is close to the Wigner surmise (see Fig. 4 (a)),
and
when $L=50$, $W=1.0$ and
the center of the energy window is set $E=3.75$,
the localization length is within the system size
and
the Poissonian behaviour is observed (see Fig. 4 (b)).

In summary, we have studied
random Dirac fermions numerically
beyond the zero modes.
Although it belongs to the orthogonal ensemble,
the zero-energy states
do not localize
and becomes critical.
The density of states $\rho(E)$ vanishes as
$\sim E^{\alpha}$ near zero energy and
the exponent $\alpha$
depends on strength of the randomness $W$.
It implies that scaling dimensions of the operators
change with strength of the randomness.
It is similar to the case of
the random gauge-field critical line found in \cite{ludwig}.
We consider that
the existence of the symmetry
${\{}{\cal H},{\gamma}{\}}=0$
is crucial to have the criticality of the zero modes.
We have studied
nature of the eigenstates
using the dimensionless conductance (Thouless number)  $g(E)$
and
the level spacing distribution $P(s)$.
As is suggested by the numerical results for the $g(E)$,
the localization length grows near zero energy
so rapidly that it exceeds the available system size
and the observables at non-zero energies exhibits
anomalous behaviour, a critical slowing down.
The $P(s)$'s near zero energy
deviate from both the Wigner surmise and the Poissonian
and exhibit critical behaviour,
as in the case of the quantum Hall states,
where rapid growth of the localization length
and the critical behaviour of the $P(s)$ were observed
\cite{ando,qhe-ls}.
We consider that it reflects the existence of critical states
at zero energy for our model.
As is discussed in \cite{ludwig},
the random guage-field critical line is unstable
in contrast to
the critical line of (1+1)-dimensional free bosons,
Tomonaga-Luttinger liquid,
and a global renormalization-group flow
for random Dirac fermions
is not fully
understood yet.
Our results show that critical states can appear
in a lattice model
and
we consider that
our study gives a clue to understand
a global phase diagram for
random Dirac fermions on a lattice.

\begin{figure}

Fig. 1
The density of states $\rho (E)$,
where $L=50$ and $W=0.7$, $0.8$, $0.9$ and $1.0$.
We have fitted the data by the power-law form
$\rho(E)=CE^{\alpha (W)}$,
where
$\alpha (W)=0.90$, $0.74$, $0.55$ and $0.39$
for $W=0.7$, $0.8$, $0.9$ and $1.0$, respectively.
We confirmed that the finite-size effect is small,
comapring the results for $L=50$
with those
for $L=30$
and $40$.

Fig. 2
(a)
The dimensionless conductance (Thouless number) $g(E)$,
where $W=1.0$ and $L=30$, $40$ and $50$.
(b)
$1/L-g(E,L)$ plot for $E=0.30$, $0.42$, $0.50$, $0.62$ and $0.70$.
It suggests a power-law form $g(E,L)\propto 1/L^{\gamma}$
rather than an exponential form $g(E,L)\propto \exp(-L/{\xi})$
in the present system size.
We consider that
it is a {\it critical slowing down}
in the available finite-size system
and
a crossover
from the power-law form to an exponential form
occurs
when the system size increases beyond the localization length.

Fig. 3
The level spacing distribution $P(s)$ near zero energy,
where
$L=50$ and ensemble average is performed
within an energy window $[0.1, 0.5]$.
We confirmed that
the finite-size effect is small,
comapring the results for $L=50$
with those
for $L=20$, $30$ and $40$.
We also found that
there is no substantial difference
between results
with energy windows
$[0.1,0.3]$, $[0.2,0.4]$ and $[0.3,0.5]$.

Fig. 4
The level spacing $P(s)$,
(a)
where $L=20$, $W=0.5$ and ensemble average is performed
within an energy window $[2.0, 3.0]$;
(b)
where $L=50$, $W=1.0$ and ensemble average is performed
within an energy window $[3.5, 4.0]$.

\end{figure}


\begin{references}

\bibitem{wu1}
X.~G.~Wen and Y.~S.~Wu,
{\it Phys.~Rev.~Lett.} {\bf 70}, 1501 (1993).

\bibitem{wu2}
W.~Chen, M.~P.~A.~Fisher and Y.~S.~Wu,
{\it Phys.~Rev.~B} {\bf 48}, 13749 (1993).

\bibitem{nnn}
Y.~Hatsugai and M.~Kohmoto,
{\it Phys.~Rev.~B} {\bf 42}, 8282 (1990).

\bibitem{graphite}
G.~W.~Semenoff,
{\it Phys.~Rev.~Lett.} {\bf 53}, 2449 (1984).

\bibitem{pi-flux}
I.~Affleck and J.~B.~Marston,
{\it Phys.~Rev.~B} {\bf 37}, 3774 (1988).

\bibitem{d-wave}
P.~A.~Lee,
{\it Phys.~Rev.~Lett.} {\bf 71}, 1887 (1993).

\bibitem{ludwig}
A.~Ludwig, M.~Fisher, R.~Shankar and G.~Grinstein,
{\it Phys.~Rev.~B} {\bf 50}, 7526 (1994);
C.~Chamon, C~Mudry and X.~G.~Wen,
{\it Phys.~Rev.~B} {\bf 53}, 7638 (1996).

\bibitem{lan}
Y.~Hatsugai, X.~G.~Wen and M.~Kohmoto,
to be published in {\it Phys.~Rev.~B}.

\bibitem{site1}
M.~P.~A.~Fisher and E.~Fradkin,
{\it Nucl.~Phys.} {\bf B251}[FS13], 457 (1985).

\bibitem{site2}
Y.~Hatsugai and P.~A.~Lee,
{\it Phys.~Rev.~B} {\bf 48}, 4204 (1993).

\bibitem{dos1}
A.~A.~Nersesyan, A.~M.~Tsvelik and F.~Wegner,
{\it Phys.~Rev.~Lett.} {\bf 72}, 2628 (1994).

\bibitem{dos2}
K.~Ziegler, M.~H.~Hettler and P.~J.~Hirschfeld,
{\it Phys.~Rev.~Lett.} {\bf 77}, 3013 (1996).

\bibitem{diracsigma}
S.~Deser, R.~Jackiw and S.~Templeton,
{\it Ann.~Phys.~{(N.Y.)}} {\bf 140}, 372 (1982);
A.~Niemi and G.~W.~Semenoff,
{\it Phys.~Rev.~Lett.} {\bf 51}, 2077 (1983);
N.~Redlich,
{\it Phys.~Rev.~D} {\bf 29}, 2366 (1984).

\bibitem{aalr}
E.~Abrahams, P.~W.~Anderson, D.~C.~Licciardello and
T.~V.~Ramakrishnan,
{\it Phys.~Rev.~Lett.} {\bf 42}, 673 (1979).

\bibitem{qhe}
See, for a review,
{\it The Quantum Hall Effect},
edited by R.~E.~Prange and S.~M.~Girvin
(Springer, New York, 1990).

\bibitem{critical}
V.~E.~Kratsov, I.~V.~Lerner, B.~L.~Altshuler and A.~G.~Aronov,
{\it Phys.~Rev.~Lett.} {\bf 72}, 888 (1994).

\bibitem{ando}
T.~Ando,
{\it J.~Phys.~Soc.~Jpn} {\bf 52}, 1740 (1983);
{\it J.~Phys.~Soc.~Jpn} {\bf 53}, 3101 (1984);
{\it J.~Phys.~Soc.~Jpn} {\bf 53}, 3126 (1984).

\bibitem{3d}
B.~I.~Shklovskii, B.~Shapiro, B.~R.~Sears, P.~Labrianides,
and H.~B.~Shore,
{\it Phys.~Rev.~B} {\bf 47}, 11487 (1993);
S.~N.~Evangelou,
{\it Phys.~Rev.~B} {\bf 49}, 16805 (1994).

\bibitem{qhe-ls}
B.~Huckstein and L.~Schweitzer,
{\it Phys.~Rev.~Lett.} {\bf 72}, 713 (1994);
Y.~Ono and T.~Ohtsuki,
{\it J.~Phys.~Soc.~Jpn} {\bf 62}, 3813 (1993);
Y.~Avishai, Y.~Hatsugai and M.~Kohmoto,
{\it Phys.~Rev.~B} {\bf 51}, 13419 (1995).

\bibitem{iiwake}
We have also calculated $P(s)$'s
for small energy windows near zero energy.
We could not obtain systematic differences among them.
Therefore we chose
a rather wide energy window
to obtain reliable statistics.


\end{references}
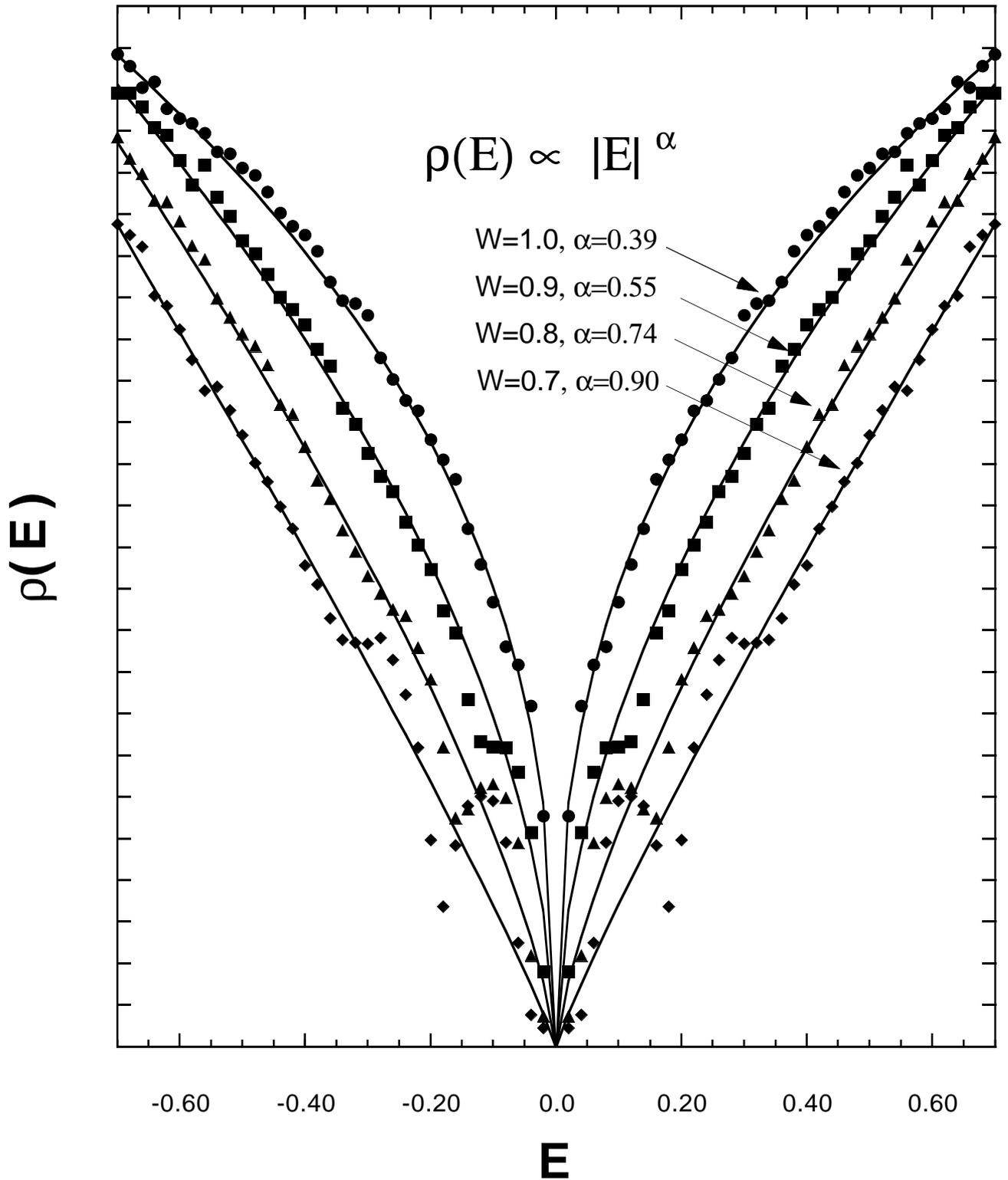
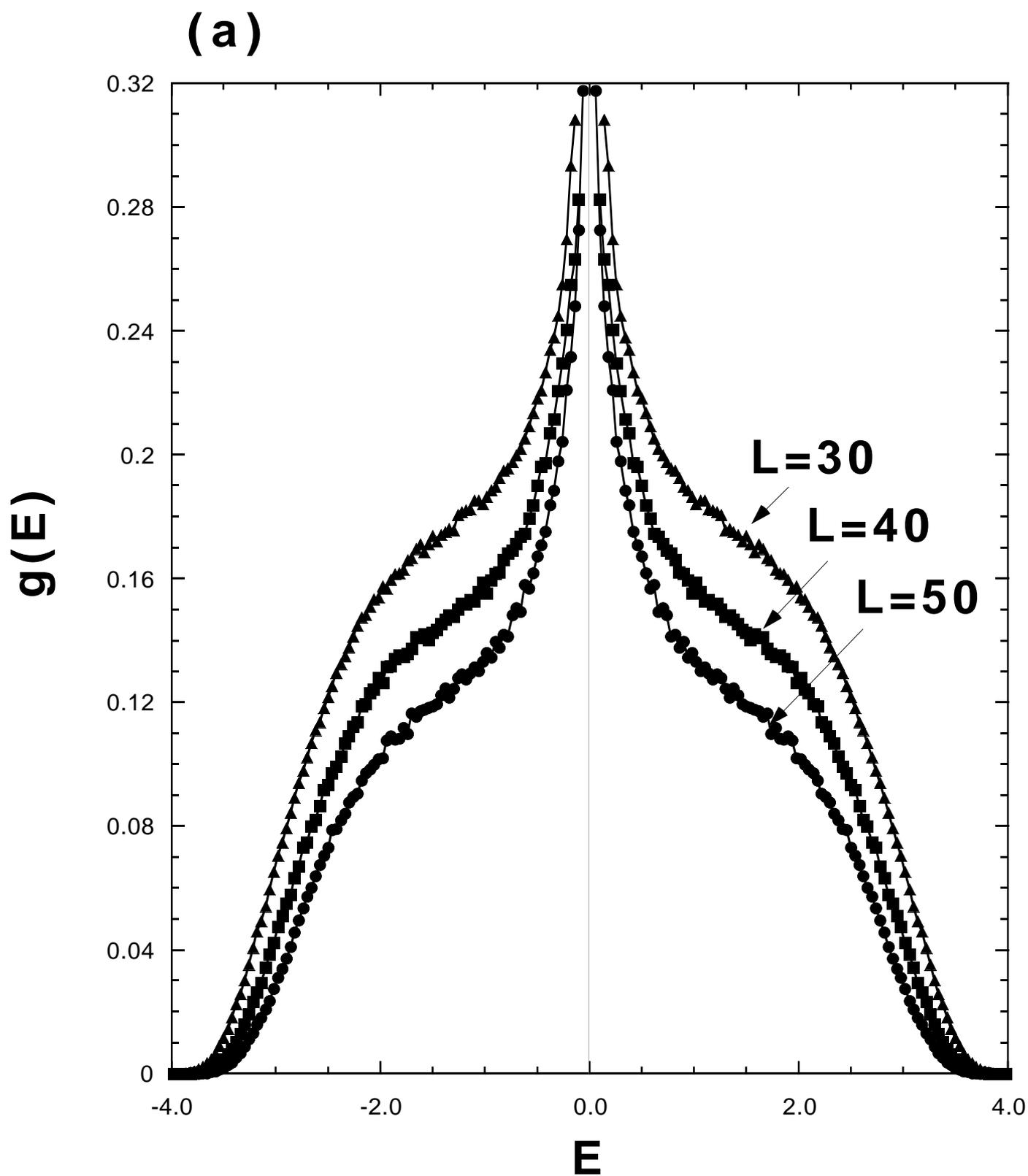
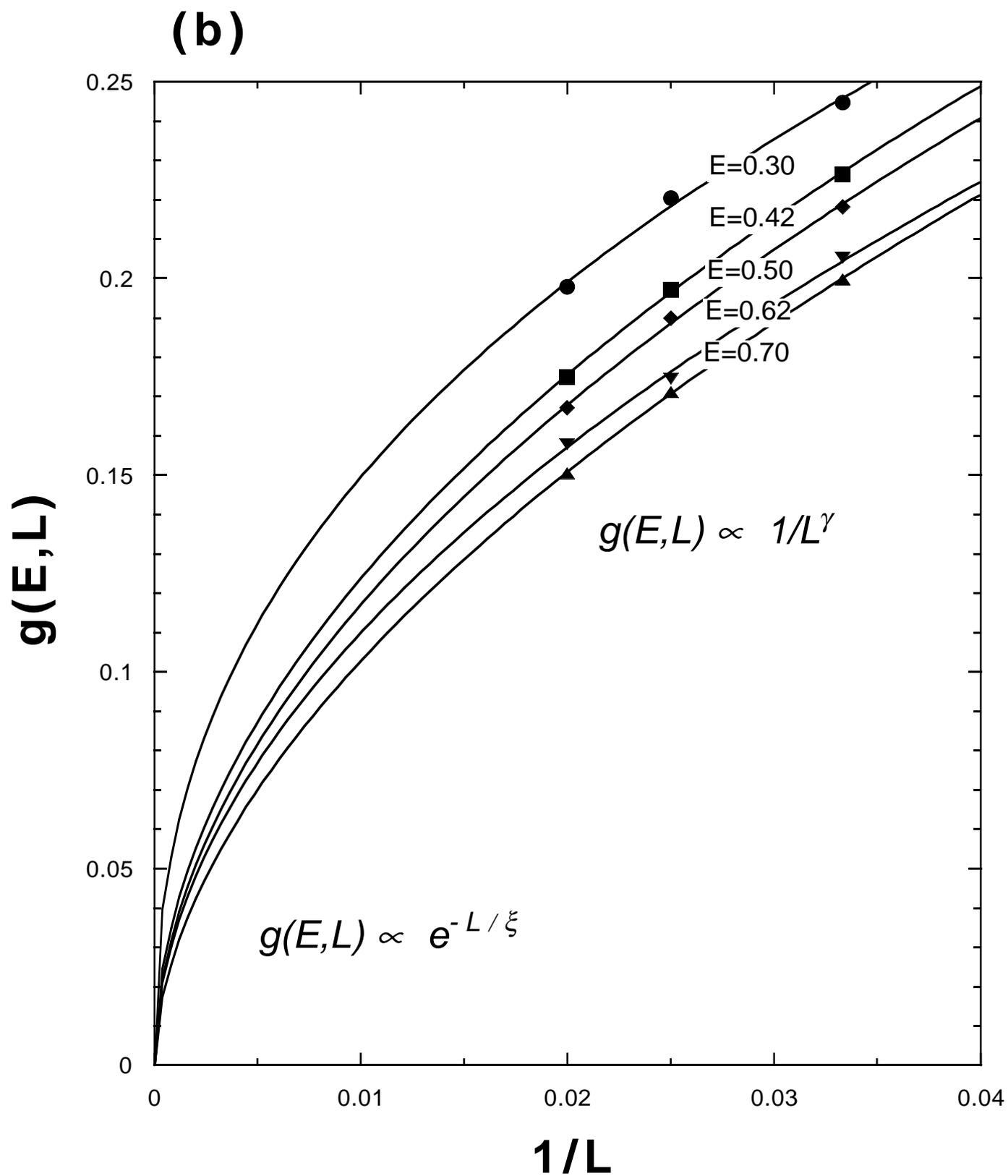
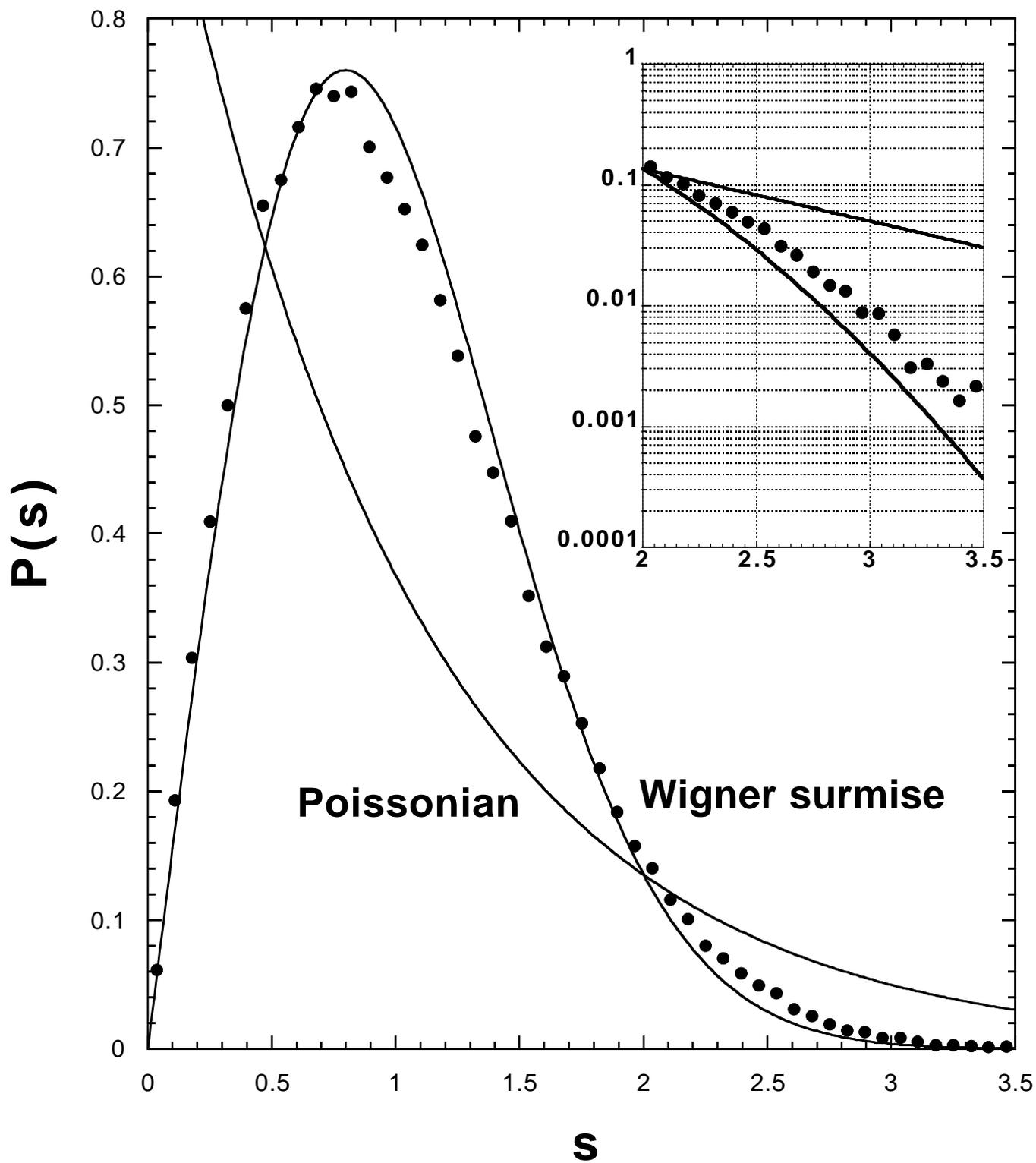
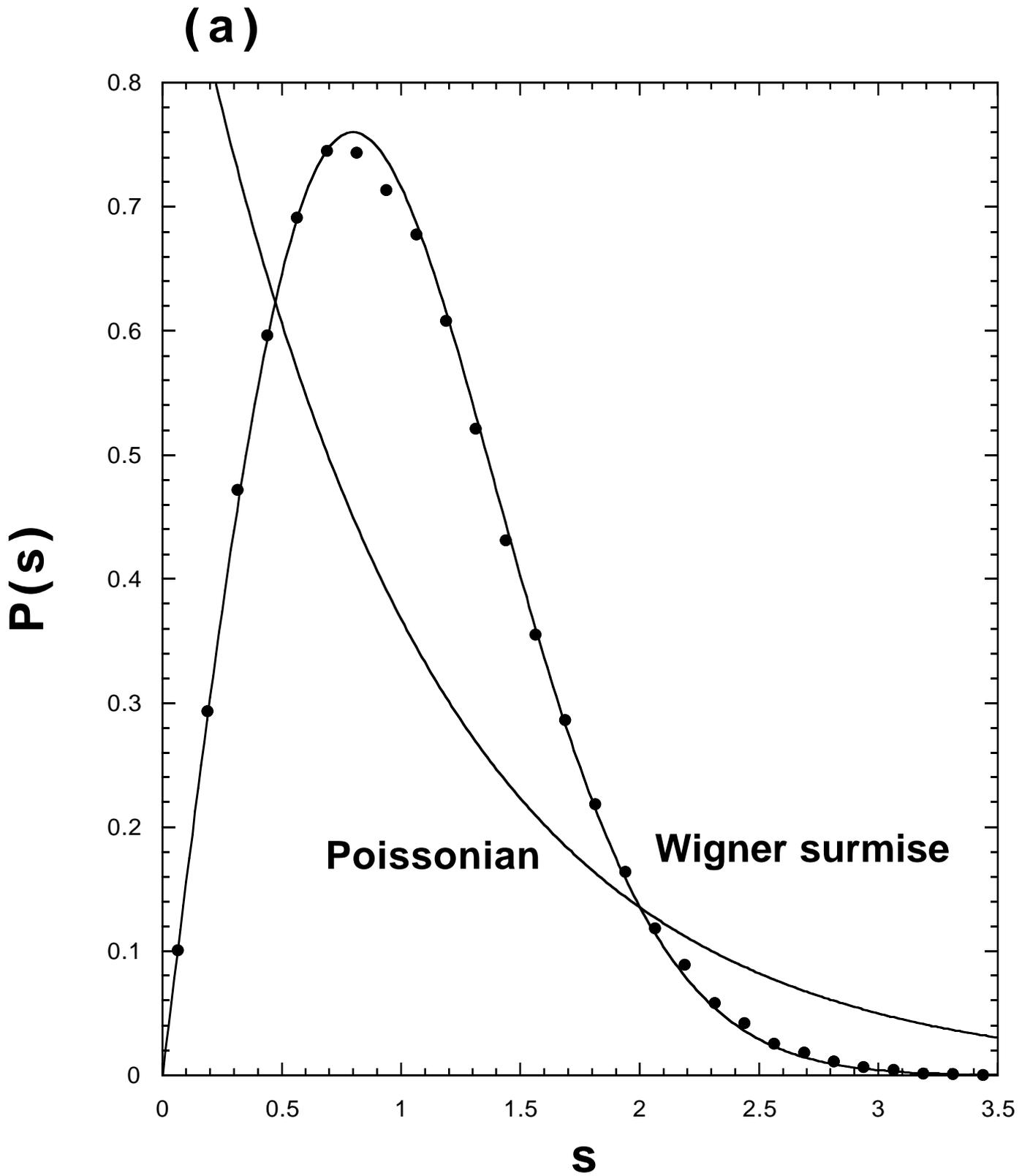
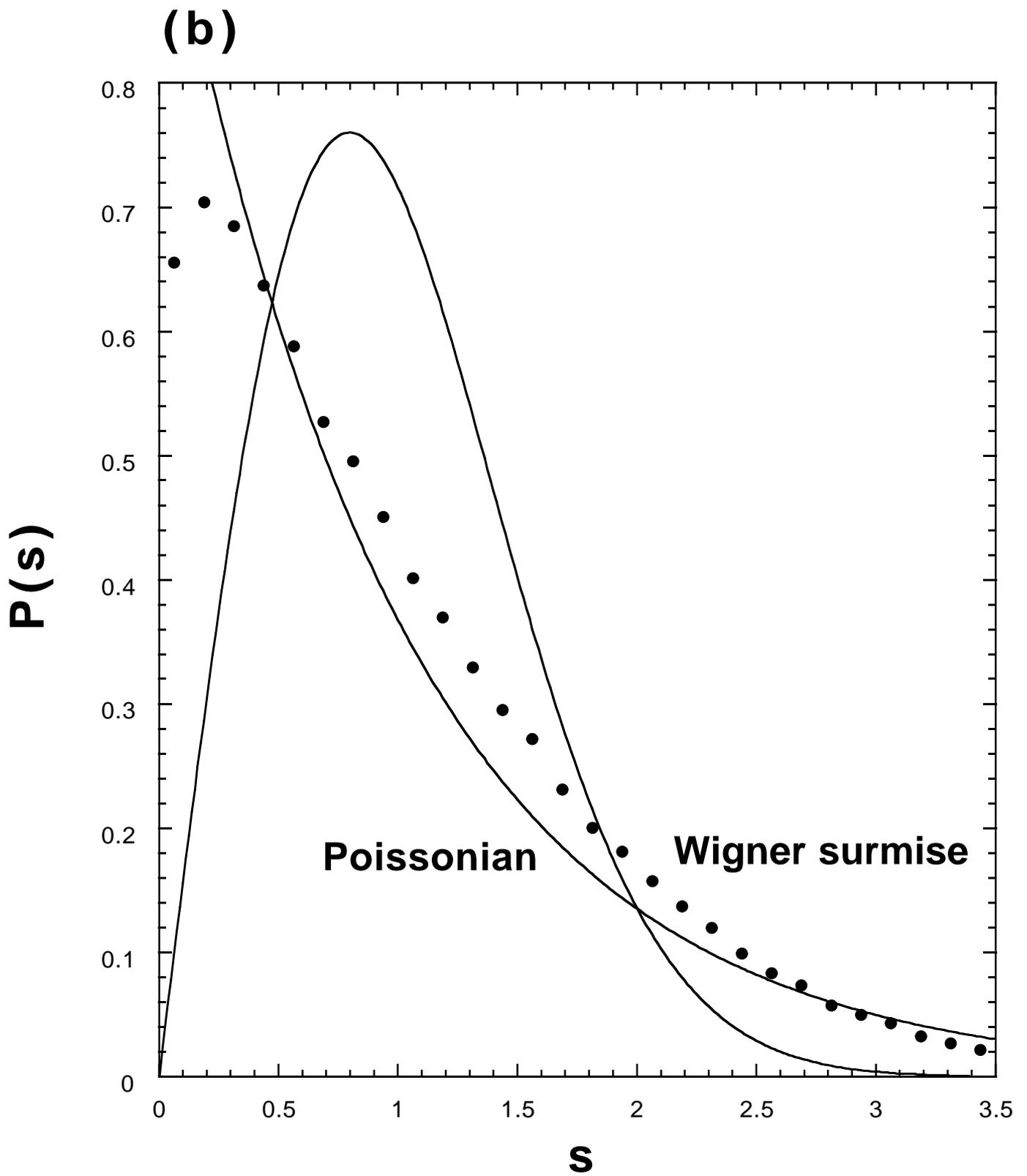
\end{document}